\documentclass{aa}
\usepackage{graphicx}
\usepackage{natbib}
\def\dR {\hbox{d$R$}}

\def\dt {\hbox{d$\theta$}}

\def\kms {\hbox{kms$^{-1}$}}
\def\eq#1{\begin{equation} #1 \end{equation}}

\begin{document}
   \title{A general catalogue of 6.7\,GHz methanol masers II:\\ statistical
   analysis}

   \author{M. R. Pestalozzi \inst{1}
          \and
          A. Chrysostomou\inst{1}
          \and
 	  J. L. Collett\inst{1}
	  \and
          V. Minier \inst{2}
          \and
          J. Conway \inst{3}
          \and
          R. S. Booth \inst{3,4}
          }

   \offprints{Michele Pestalozzi \\
email: michele@star.herts.ac.uk \\}

   \institute{Centre for Astrophysics Research, University of
   Hertfordshire, College Lane, AL10 9AB Hatfield, UK,\\ 
             \email{M.R.Pestalozzi,a.chrysostomou,j.l.collett@herts.ac.uk}
                     \and
              Service d'Astrophysique, DAPNIA/DSM/CEA Saclay,
              91 191 Gif-sur-Yvette, France, 
              \email{Vincent.Minier@cea.fr}
                     \and
              Onsala Space Observatory, 439 92 Onsala, Sweden, 
              \email{jconway,roy@oso.chalmers.se}
	             \and
              Hartebeesthoek Radio Astronomy Observatory, PO Box 443,
   Krugersdorp 1740, South Africa 
           }
   \date{Accepted 24 November 24 2006}

\authorrunning{Pestalozzi et al.}
\titlerunning{Maser catalogue, the statistics}

   \abstract{Methanol masers at 6.7\,GHz are recognised markers of high-mass
   star formation regions. The study of their distribution in the Galaxy gives
   important insights into the star formation activity of the Milky Way. We
   present a statistical analysis on the General Catalogue of 6.7\,GHz
   methanol masers in the Galaxy with the aim of extracting global properties
   of the masers.}{We provide constraints
   on the luminosity  
   function of 6.7\,GHz methanol masers and on their total number in
   the Galaxy.}{We model the spatial distribution of the masers in the Milky
   Way by using their distribution in galactocentric distance which is
   unambiguous once a rotation curve for the Galaxy is assumed. This
   is the starting point for determining the luminosity function
   of the masers.}{The
   luminosity function of 6.7\,GHz methanol masers is modelled as a power-law
   with sharp cutoffs and having an index lying between
   $-1.5$ and $-2$. We also predict the
   number of detections of methanol masers assuming different
   sensitivity limits in the observations.}{}


   \maketitle
%

\section{Introduction}
\label{sec:intro}

Since its discovery by \citet{bat87} and \citet{men91a} the bright
maser emission from methanol (at 12.2 and 6.7\,GHz respectively) has
become a reliable tool for detecting and studying regions where 
(massive) stars form and are in their very early stages of
evolution (see e.g. \citealt{elli06}). Methanol masers were divided
into two empirical classes, I and II \citep{men91a}. Class II 
  masers are detected close to strong Infrared sources (as e.g. Ultra Compact
  (UC) \textsc{Hii} regions), while class I are observed offset from these
  objects, in the shock regions of their outflows. Theoretical modelling was
  able to identify class I masers as collisionally pumped masers, while class
  II as radiatively pumped ones (e.g. \citealt{sob97,cra01,cra05}). 

Observations of methanol maser sites at other wavelengths support the
paradigm of methanol masers as one of the first signposts of massive
star formation (e.g. \citealt{wal99,goe02,pes02b}). Being very
bright, methanol masers are ideal for high spatial resolution observations
using interferometers where very detailed positioning as well as mapping
of the finest spatial and dynamical maser features gives important insights
into the nature of some known sources (e.g. \citealt{min01,min03,pes04a}).
Nevertheless, the question of whether these masers trace discs or outflows
in young protostars still remains open to debate (e.g. \citealt{deb03}).

Their association with youthful and massive star birth opens up the
possibility of using methanol masers as a new and reliable tracer of that
rapid stage of evolution prior to the development of UC 
\textsc{Hii} regions. For more than a decade now, searches for new
methanol maser sources have been undertaken by a number of authors, e.g.
\citet{mcl92a,schu93,cas95a,elli96,szy00a,szy02,pes02a}. All together, this
brought the number of known methanol masers to 519 \citep{pes05}. This
represents a statistically significant sample which, for this paper, 
motivates us to study their spatial distribution throughout the Galaxy
with the aim of determining their luminosity function. 

To date, most statistical work has focused on finding correlations between the
physical characteristics of the maser and the associated IRAS source. In
general, 6.7\,GHz methanol masers seem to be more efficiently detected
towards bright IRAS sources, having $F_{60} > 100$~Jy
\citep{vdw96,szy00b}, with a clearly higher detection rate in the inner
than in the outer Galaxy \citep{szy00b}. Early attempts to find a
correlation between the associated IRAS flux density with the maser flux
density gave no positive results. In \citet{vdw95} the authors are left
  with the fact that the maser flux densities are smaller than the 100$\mu$m
  ones, suggesting that the masers could be pumped by the 100$\mu$m
  photons. The lack of correlation between maser flux density and IR flux
  density is not explained in that work. IRAS 
sources associated to 6.7\,GHz methanol masers seem to be concentrated in
a small region of the [25-12]--[60-25] colour-colour diagram, indicating
that the maser arises in specific environments \citep{szy00b}. Based on
the same conclusions, \citet{xu03} claim that the methanol maser phase
must be short and occurs in the early stages of star formation, where the IR
luminosity is high enough that the IR radiation itself might be
responsible for the maser pumping, in accordance with theoretical
modelling \citep{sob97,cra05}. One should note, however, that the
association of methanol masers with IRAS sources is not a reliable one, as
several studies at high resolution have confirmed (e.g.
\citealt{elli96,min02a,pes02a}): methanol masers are often seen offset from all
IRAS or centimetre continuum sources. 

The range of velocities and the linewidths of methanol masers have also
been used to try to deduce some intrinsic characteristics for these
objects. For instance, \citet{sly99} found that methanol masers show a
clear velocity difference relative to the velocity of the parent cloud (a
comparison made with the velocity of CS). This could suggest an origin for
methanol masers 
in discs seen edge-on. Alternatively, \citet{szy00b} suggested that the
velocity dispersion seen in maser spectra could be used as an evolutionary
tool, arguing that the early stages of the formation of an UC
\textsc{Hii} region would be traced by narrow line masers because 
of the low number of masing clouds and the lower velocity dispersion
within them. Once the UCHII region begins to disperse the surrounding
circumstellar matter, larger linewidths are expected. All such hypotheses
which appeal to linewidths or the radial velocities of spectral features
are subject to the caveat that spectral feature blending could falsify the
measurement of the linewidth resulting in, for example, a false 
evolutionary sequence.

Comparisons have also been made between masers at different frequencies.
\citet{sly99} find that in those sources where 6.7 and 44\,GHz masers
coexist, the ratio of their flux density can be used to discriminate class
I from class II masers. On the
other hand, no correlation was found when comparing 6.7 with 12.2\,GHz
counterparts, except for the clear fact that 6.7\,GHz masers are always
found to be stronger in flux and in brightness temperature when assuming a
constant source size \citep{mal03}.

More recent studies have concentrated their efforts in characterising
the sources hosting methanol masers
(e.g. \citealt{hil05,pur06}). Methanol masers seem to be always
associated with 1.2\,millimetre dust continuum emission as well as
CH$_3$CN emission. Maser-hosting sources are in general more massive
and hotter than sources showing 1.2\,mm continuum emission but no
maser. Also, sources hosting methanol maser show hot core
characteristics and most of them pass the Lumsden MSX criterion for
massive protostars in the Galactic plane \citep{lum02}. Finally,
methanol masers have been detected toward dark mid-IR clouds,
clearly indicating that star formation is ongoing in those regions. 

The luminosity function of 6.7\,GHz methanol masers has been studied using
different approaches. The
main problem is in determining the heliocentric 
distance for every source. Because of the high correspondence of 6.7\,GHz
methanol masers with OH masers, \citet{cas95a} assume that the
luminosity function of 6.7\,GHz methanol masers should be similar to
the one for OH masers (see \citealt{cas87b}) from which they conclude
that there must be some 500 methanol maser sources in the Galaxy. In
\citet{vdw96}, the ambiguity of 
kinematic distances is solved in a probabilistic way. When the
decision was not clearly made on the basis of
the total luminosity of the IRAS source hosting the maser, the source
was assigned a probability for it to lie at the near heliocentric
distance. Repeating this assignment in a random way (and for different
probabilities) and averaging over
the results the authors obtained a spatial distribution from which the
luminosity function was estimated. For their sample of about 240 sources, the
authors fit a power-law luminosity function with an index of $\approx
-$2 (the fitting was done on the distribution of sources per unit
luminosity interval). This index is expected to flatten
in the low-luminosity end of the distribution as otherwise the total
number of methanol masers in the Galaxy would be far too
high. \citet{elli96} refrained from any luminosity function estimate,
invoking the problem of determining the distance to every source.  
More recent studies of large samples of 6.7\,GHz methanol masers by
\citet{szy00b,szy02} have not directly addressed the study of the
luminosity function of methanol masers.

The most recent 
statistical study of methanol masers in the Galaxy has been presented by
\citet{vdw05} in which estimates for the lifetime of the methanol maser
phase and the total number of methanol masers are made.
The model uses a combination of the initial mass function, (local) star
formation rate, a synthetic distribution of these sources in the
Galaxy and a constant detection limit of 1.5\,Jy in some template
  regions to 
conclude that methanol masers should last between $2.5-4.5\times
10^{4}$~years and that there should be around 1100 in the Milky Way.

The aim of this paper is to present a statistical analysis of the
population of 6.7\,GHz methanol masers in the Milky Way. We aim to
constrain the luminosity function, and begin by modelling the spatial
distribution of methanol masers in our Galaxy. This approach is
 different from the one presented in \citet{vdw96} and \citet{vdw05}, where
the authors apply a {\it random choice} algorithm to solve the
heliocentric distance ambiguity for sources on galactic orbits
internal to the Sun. Our starting point is the distribution of
methanol masers in (kinematic) galactocentric distance, which is unique once a
rotation curve for the Galaxy is assumed. We then assume that that 
distribution is the azimuthally averaged surface density of sources in the
Galaxy. In this way we are not compelled to solve the 
distance ambiguity for every source lying on galactic orbits internal
to the Sun, as the decision is made by the surface density itself. 

Nevertheless, the estimate of the total number of sources in
\citet{vdw05} will be used as check for consistency in this
work. Furthermore, the present study represents a study of principles
with fewer assumptions than \citet{vdw05}. Finally, the physical
meaning of the luminosity function of the 
maser emission is not addressed in this paper.


\section{Modelling the maser distribution and luminosity function}
\label{sec:stat}

The determination of the luminosity function for astronomical sources
requires an accurate knowledge of their distance from the observer. In the
case of galactic sources, kinematic distances from spectral line
observations are becoming more accurate thanks to improved modelling of the
Galactic rotation curve (e.g. \citealt{bra93} as well as \citealt{rus03}).
The difficulty resides in the discrimination between {\it near-} and {\it
far}- heliocentric distances for all sources on orbits {\it internal} to
the Solar orbit around the Galactic Centre. 

\begin{figure*}[!th]
\begin{center}
\includegraphics[angle=90,width=18cm]{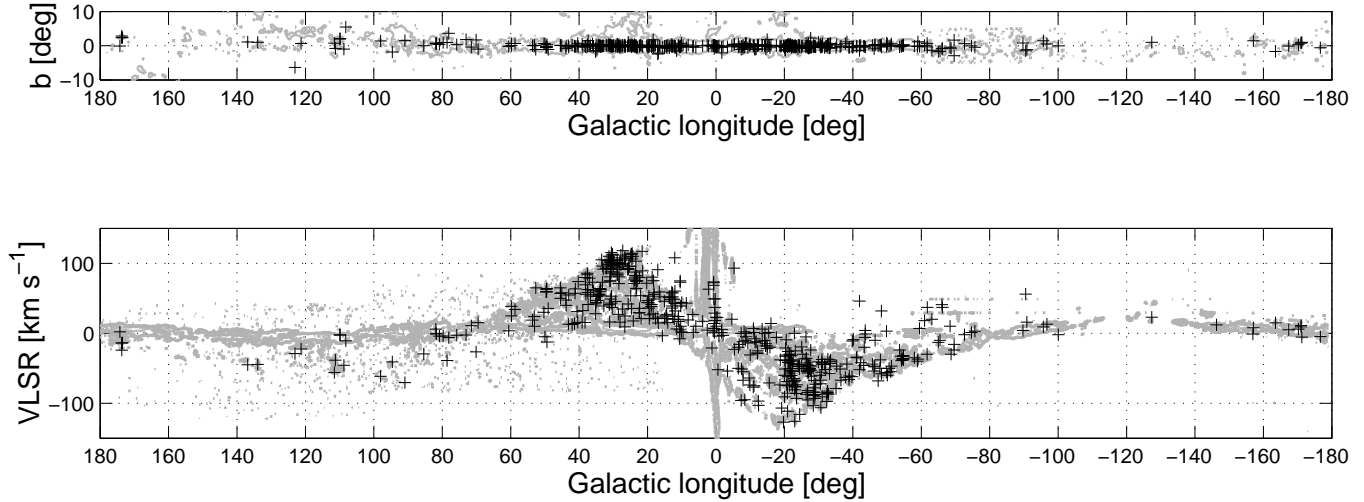}
\caption{Distribution of methanol masers in Galaxy, superposed on the CO
  contours from \citet{dam87}, in space (top) and LOS velocity
  (bottom). The methanol masers seem to accurately follow the overall
  structure of the Galaxy, both in space and LOS
  velocity. Particularly visible in the bottom panel is the fact that
  methanol masers are tracing the spiral arms ($150^{\circ} > l >
  80^{\circ}$ and $-40^{\circ} > l > -90^{\circ}$) and the high
  rotational velocity of the nuclear ring ($l \approx 0^{\circ}$).}  
\label{fig:meth+co}
\end{center}
\end{figure*}

Beside the probabilistic approach presented in \citet{vdw96}, another
  way to solve the heliocentric distance ambiguity is the observation
  of \textsc{Hi} self-absorption. The
method relies on the fact that molecular clouds have an outer layer of neutral
hydrogen which can absorb the ubiquitous background galactic \textsc{Hi}
  emission. Sources lying on the near  
side of the galactic centre would have more \textsc{Hi} emission to
  absorb than sources 
on the far side, making it straightforward to solve the ambiguity
  \citep{jac02,lis81}. As the previous one, this method still relies
  on the calculation of kinematic 
distances, which requires a good knowledge of the rotation curve of the
Galaxy. Recently, \citet{bus06} used this method to successfully
  determine the distances of a number of massive young stellar object
  candidates in the fourth quadrant of our 
  Galaxy. The authors succeeded to resolve the distance ambiguity
  for 80\% of their targets, 
  the main limitations being the coarse spatial resolution of the 
  archival \textsc{Hi} data as well as the inability to apply the method for
  sources at galactic longitudes $\ell < \,\mid 15^{\circ} \mid$.

The main idea of the present study is to avoid solving the distance ambiguity
for every single source and adopt a global approach instead. This approach
starts with a model of the spatial distribution of methanol
masers in the Galaxy. We then model the observability of sources with different
luminosities at different detection limits and estimate their real total
number.  

The sample of methanol masers used for the present study is the General
Catalogue of 6.7\,GHz Methanol Masers (GCMM) published by \citet{pes05},
available on the CDS website\footnote{http://vizier.cfa.harvard.edu/viz-bin/VizieR?-source=J/A+A/432/737/}.
This sample is far from complete, but allows a good study of
  principles and a first test for the formalism we adopt. The outcomes of this
study will be compared to the ones from the Methanol MultiBeam Survey
(MMB\,Survey)\footnote{The MMB Survey aims at surveying a
  strip of $b= \mid 2^{\circ}\mid$ across the galactic plane and at all
  longitudes, searching for
6.7\,GHz methanol masers. It has started in January 2006. See also
http://www.jb.man.ac.uk/research/methanol/ }. The
MMB Survey is expected to  
probe the whole Galaxy at a depth 
  of 0.1\,Jy at 1$\sigma$, and will probably reveal the complete population of
  6.7\,GHz methanol masers in the Galaxy.

\subsection{Spatial distribution of the masers}
\label{sec:spadistr}

For a given observational sensitivity, the {\it observed} spatial
distribution of a certain type of object in the Galaxy is strongly dependent
on the luminosity function of those objects. If, for instance, the
luminosity function is strongly peaked towards low luminosities, all 
searches for those objects at a certain detection limit will select the
majority of sources in the solar neighbourhood and miss those lying further
away. This will give an observed spatial distribution of sources which will
probably not reflect the real distribution of sources. If on the other hand
the spatial distribution is known to a high degree of confidence it is then
possible to estimate both the luminosity function as well as the {\it real}
total number of those sources. 

A general view of the methanol masers known to date superimposed on
the CO emission in the galactic plane both in space and LOS velocity
is presented in Fig.~\ref{fig:meth+co}. The masers clearly follow both
the structural and dynamical features of the CO gas in the Milky
Way. There is a clear concentration of sources at longitudes close to $\pm
50^{\circ}$ which also follows the dynamical signature of the
molecular ring. Also, the masers seem to be concentrated in the
galactic plane (see also \citealt{pes05}). Finally, there are some
masers that seem to mark the 
nuclear ring ($l \approx 0^{\circ}$), and hence lie within 1\,kpc from
the Galactic Centre. This is not the first time that masers are
detected within 1\,kpc of the Galactic Centre (see e.g. the detection
of OH masers, \citealt{cas83}). The discussion on this point goes
beyond the scope of this paper, but it is nevertheless interesting to note
that, being methanol masers exclusively associated with star formation
activity, these detections are strong indications of recent star
formation very close to the Galactic Centre.

\begin{figure}
\begin{center}
\includegraphics[width=9cm]{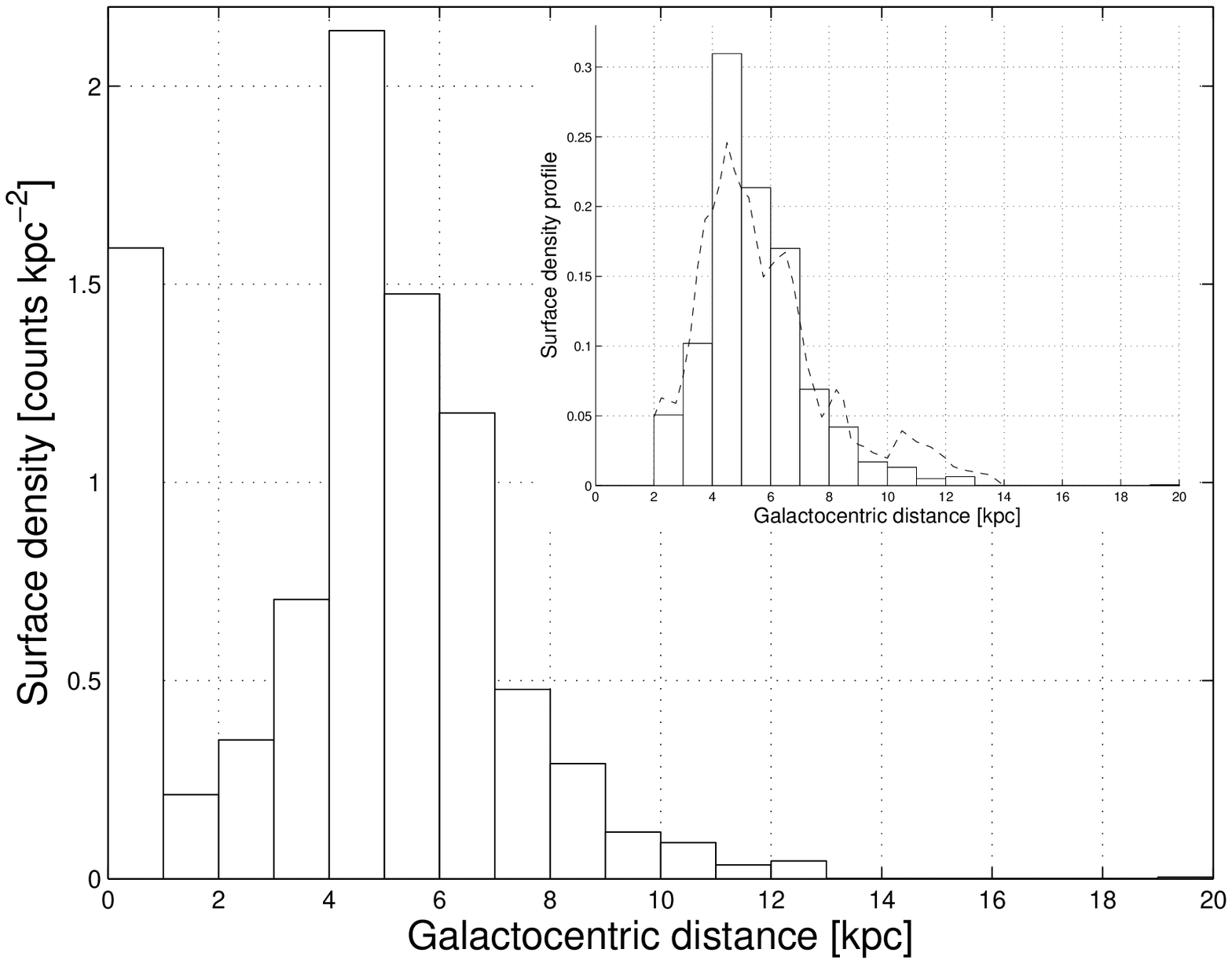}
\caption{{\it Main panel}: Spatial density of methanol masers in the
  Galaxy as a 
  function of galactocentric distance. {\it Inset}: normalised profile
  of the surface density of the masers (histogram) compared with the
  surface density profile of the H$_2$ gas (dashed line, from
  \citealt{bli96}). See Table 
  \ref{tab:res_spafit} for the results of the fits to the histogram in
  the main panel (excluding the first bin).} 
\label{fig:n_gc}
\end{center}
\end{figure}

The starting point in our
analysis is the distribution of masers against their galactocentric
distance, as this does not suffer from any distance ambiguity. The
distance of every source from the galactic 
centre is uniquely determined by its galactic longitude and LOS velocity, once
a rotation curve for the Galaxy is assumed (we adopt the rotation
curve of \citealt{bra93})\footnote{Note that for the calculation of
  the galactocentric distance we use the line--of--sight velocity of
  the brightest spectral component. As mentioned in \citet{pes05}, the error
  of that calculation is about 1\,kpc when taking into account the
  inaccuracies in the position and in LOS velocity.}. Figure
\ref{fig:n_gc} shows the surface density of methanol masers in the
Galaxy (main panel) and the surface densities of methanol
masers and molecular gas (inset, histogram and dashed line respectively)
normalised to the integral under the curves. The ratios 
between the values at 5\,kpc (peak, or molecular ring) and
the values around the Sun ($\approx$ 8-9\,kpc) are 6:1 and 5:1 for
masers and gas respectively. Knowing that methanol masers 
are associated exclusively with star formation regions, this fact
could support the idea that star formation is more efficient in the
molecular ring than in the outer Galaxy. We do not draw this
conclusion, as this effect could be due to a lower maser detection
rate coming from an uneven observational coverage of the outer
Galaxy. Another interesting empirical fact is given by the ratio of
  the gas surface density to the maser surface density. This ratio 
  is an estimate of solar masses of gas per methanol masers and it has
  values between $1.5\times 10^{6}$ and
$9.7\times 10^{7}$ for peak and solar neighbourhood,
  respectively. These values are 
comparable to the gas content in large molecular clouds ($\approx
10^{7}$M$_{\odot}$, see e.g. \citealt{kna93}), which means that in the ring we
are expecting $\approx$\,10 methanol masers per large molecular cloud,
while in the outer Galaxy this number drops to approximately one maser
every 10 clouds. In the same line of thought we can do an order of
magnitude comparison of these numbers with the
total content of gas in the Milky Way ($\approx 10^{9}$M$_{\odot}$,
see e.g. \citealt{dam93}) and obtain an empirical estimate of the
total number of methanol masers in our Galaxy, which should be of the
order of $10^{3}$ sources.

We consider the histogram in the main panel of Fig.~\ref{fig:n_gc} to be a
sufficiently reliable signature for the shape of the spatial
distribution of methanol masers in the Galaxy. We support our
assumption with the fact that the masers seem to accurately follow the
distribution of CO in the Milky Way both in space and LOS velocity
(Fig.~\ref{fig:meth+co}). 
We make two assumptions: that the surface density of methanol
masers in the Galaxy, F(x,y), can be projected onto the Galactic
plane ($\mid b\mid=0^{\circ}$), and that the masers are distributed
axisymmetrically, i.e. the surface density distribution depends only
on the galactocentric radius, $R$. The total number of sources
in an annulus of thickness $\dR$ is then $F(R)\,2\pi\,R\,\dR = H(R)\dR$, where
$H(R)$ is the function fitted to the
histogram in Fig.~\ref{fig:n_gc}. We support our two assumptions with the
following arguments: a) the distribution of methanol masers in galactic
latitude has a very small FWHM (about 1.0$^{\circ}$, see \citealt{pes05}); and
b) axisymmetry is assumed by noting that the clear peak in the histogram in
Fig.~\ref{fig:n_gc} is at a galactocentric distance where the spiral
pattern of the Galaxy is not yet clearly visible (molecular ring, see
e.g. \citealt{rus03}). The latter argument is also supported in 
\citet{fux99}, where it is shown that the 
existence of a bar in our Galaxy does not significantly affect the
dynamics outside the molecular ring.  The total
number of observed sources in the Galaxy $N_{tot-obs}$ is given by the 
integral from 0 to $R_{max}$ of the function $H(R)$. The Gaussian
profile used for the fitting is parametrized by the integral under the curve
$N$, the mean $R_{peak}$ and width 
$\sigma$. The results of the fit are shown in Table~\ref{tab:res_spafit}.

\subsection{Luminosity distribution of methanol masers}

\subsubsection{Step 1: equal luminosity for all masers}
\label{sec:howmany}

We use the assumption of the masers having all the same luminosity in
order to better understand the basic trends and behaviour of
the model within a simple framework.

\begin{figure}[ht!]
  \centering
  \includegraphics[width=7cm]{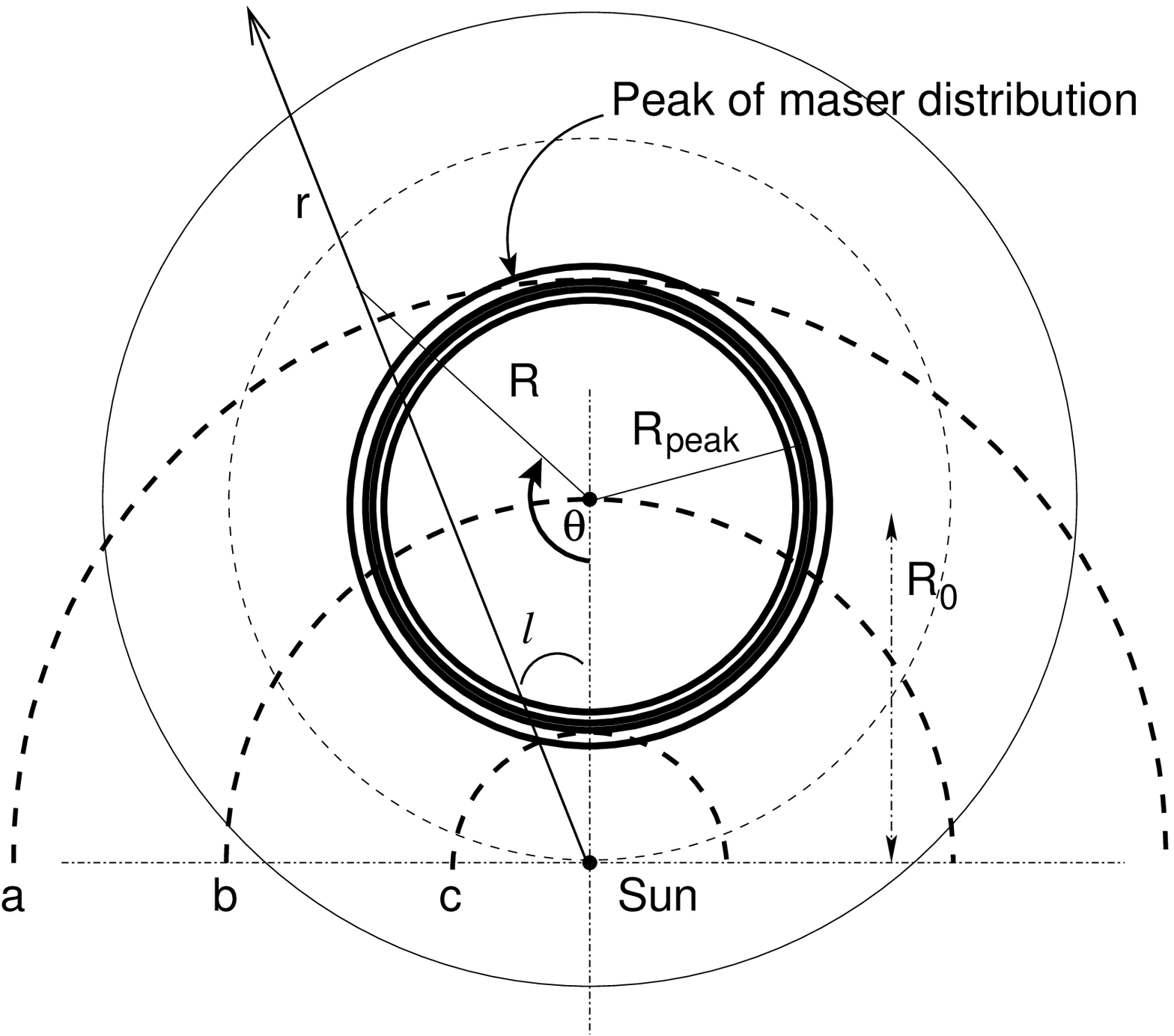}
   \caption{Geometry of the modelled methanol masers in the
  Galaxy. $R_{peak}$ is 
  the radius at which the maser distribution peaks, $R_0$ the radius of the
  solar orbit around the galactic centre, $\theta$ the central angle,
  $r$ the heliocentric distance and $\ell$ the galactic longitude. The dashed
  arcs {\bf a}, {\bf b} and {\bf c} are equidistant curves from the
  Sun, described in the text.} 
  \label{fig:distr}
\end{figure}

\begin{figure}
\begin{center}
\includegraphics[width=8cm]{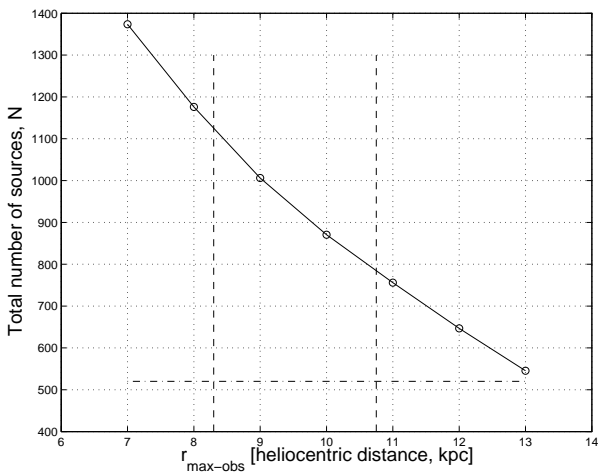}
\caption{Dependence of the real total number of sources $N$ inferred by
  fitting a distribution $F$ truncated at different values of
  $r_{max-obs}$ to the histogram in Fig. \ref{fig:n_gc}, using the
  assumption of equal luminosity for all masers. The range of $r_{max-obs}$
  defined by the vertical dashed lines corresponds to the range for $N$
  estimated in \citealt{vdw05}. The horizontal dashed line indicates
  the total number of sources in GCMM, 519.} 
\label{fig:trunc}
\end{center}
\end{figure}

If we assume that all methanol masers have the same luminosity, the
sensitivity limit of the observations translates into a maximum
heliocentric distance $r_{max-obs}$, within which masers will be
detected (dashed arcs {\bf a}, {\bf b} and  {\bf c} in
Fig.~\ref{fig:distr}). We fit H(R) to the 
histogram of Fig.~\ref{fig:n_gc} when applying different $r_{max-obs}$,
i.e. with a spatial distribution which is {\it truncated} due to
sensitivity. Recalling our basic 
assumption that the histogram in Fig.~\ref{fig:n_gc} is a reliable signature
for the shape of the galactic distribution of masers, we fix
$R_{peak}$ and $\sigma$ of the function $F(R)$ found from the original
fit to the histogram (Table \ref{tab:res_spafit}) and fit only for
$N$. This will be the total number of sources in the Galaxy. We have then:
\eq{N_{tot-obs} = \int_{-\theta_{max}(R')}^{+\theta_{max}(R')}\,
  \int_{R_{min}}^{R_{max}} F(\theta,R')\,R'\,\dR'\dt 
\label{eq:ntobs}
}
where $R_{min/max}$ are defined by the heliocentric radius
$r_{max-obs}$ and the function $\theta_{max}(R)$ ensures that masers
are only counted if they lie within $r_{max-obs}$. Note that if
$r_{max-obs} > R_0$ (i.e. the depth of the observations reaches
beyond the Galactic Centre) then $\theta_{max} = \pi$ for
$0<R<r_{max-obs}-R_0$ (arc {\bf a} in Fig. \ref{fig:distr}). 
The results of the fits are
summarised in Table~\ref{tab:res_spafit} and in
Fig.~\ref{fig:trunc}. We choose 7\,kpc as minimum value for
$r_{max-obs}$, as shorter distances would be strongly 
inconsistent with the observed distribution of masers in the
Galaxy. This is visible from Fig. \ref{fig:n_gc}, where the histogram
shows that there are sources close to the galactic centre, i.e. at
8.5\,kpc from the Sun. The fits are indicative of how many sources
there should be in the Galaxy if the sample was severely limited by
sensitivity. 

Increasing $r_{max-obs}$ to large values 
(e.g. 40\,kpc) is equivalent to stating that the 
sample in GCMM does not depend 
on the luminosity distribution of the masers, and hence that all
masers in the Milky Way are detected. There are several
reasons to believe that GCMM does not 
contain all methanol masers sources in the Galaxy: not only is there a
bias due to the unknown luminosity function but also the catalogue is 
biased due to non-uniform coverage of the galactic plane. The
present study is meant to make predictions on the basis of the data
available. 

Note that, by introducing a sensitivity limit, i.e. $r_{max-obs}$, we define a
minimum luminosity for all detected masers. For a detection limit of
1\,Jy over 0.2\,\kms{} or 5.56\,kHz at the rest frequency of
6.668519\,GHz (\citealt{mul04}), we get minimum luminosities of the 
masers of $2.85\times 10^{-6}$, $1.91\times 10^{-7}$ and $7.33\times
10^{-8}$~\,$L_{\odot}$ for $r_{max-obs}$ equal to 40.5, 10.5 and
6.5\,kpc, respectively. 

\begin{table}
\begin{center}
\caption{Results of the best fit to the histogram in
  Fig.~\ref{fig:n_gc}. The values of $r_{max-obs}$ indicate a
  truncation of the function $F$ due to sensitivity. A large value of
  $r_{max-obs}$ (as e.g. 40\,kpc) is equivalent to no
  truncation (see text for an explanation). The column $N$  
  lists the {\it real} total number of sources. The mean and 
  width for the Gaussian distribution are 5.08 and 1.42\,kpc respectively.} 
\label{tab:res_spafit}
\begin{tabular}{ccc}
\hline\hline
Profile & $r_{max-obs}$ [kpc] & $N$ \\ 
\hline
Gauss & 40 & 483 \\
 & 12 & 668 \\
 & 10 & 907 \\
 & 8 & 1228 \\
\hline
\end{tabular}
\end{center}
\end{table}

Comparing these results with recent estimates of the total
number of methanol masers in the Galaxy (as e.g. in Sect.
\ref{sec:spadistr} and \citealt{vdw05}), and retaining our assumption
that all masers have the same luminosity, we conclude that GCMM
contains sources up to the galactic centre and just beyond (see dashed
vertical lines in Fig.~\ref{fig:trunc}).

The conclusions we can draw at this point of the paper are:
\begin{itemize}
\item from fitting the histogram in Fig.~\ref{fig:n_gc} with {\it
  truncated} models of the spatial distribution (eq. \ref{eq:ntobs})
  we see that GCMM has a 
  considerable number of sources missing (up to a factor of about
  2-3);
\item this suggests that the luminosity function of 6.7\,GHz
  methanol masers is not a delta function (i.e. all masers do not have the
  same luminosity) and it is most likely dominated by intrinsically weak
  emitters, potentially young high-mass star formation regions;
\item the large number of missing sources is mostly due to the shallow
  sensitivity limit of most of the observations which contribute to
  the GCMM (see \citealt{pes05}). Only a very small part of the Galaxy 
  was observed with a sensitivity limit better than 2\,Jy.
\end{itemize}

\subsubsection{Step 2: introducing a distribution of luminosities}

The general distribution of luminosities of methanol masers is the
link between the real spatial distribution and the 
observations. The observability of a source depends on a combination of its
luminosity and the sensitivity of the survey used to detect
it. High-sensitivity observations (e.g. 0.1\,Jy at 1$\sigma$) can detect
low-luminosity sources far away from the Sun, reaching a high level of
completeness. Deep surveys guarantee that one probes most, if not
all, of the Galaxy: this is the aim of the MMB Survey. 

By discarding the assumption of equal luminosity for all masers and
introducing a luminosity distribution $G(L)$, $r_{max-obs}$ no
longer defines any sharp spatial cutoff point. The physical depth of
the observations will scale as $\sqrt{L/\Phi}$, where $L$ is
luminosity and $\Phi$ is the detection limit in the
observations. It is therefore not possible to define an absolute value
for the largest distance probed by the observations, as this will vary
within the same combination of luminosity function and sensitivity of the
observations. For every sensitivity limit $\Phi$, the scaling of the
spatial cutoff becomes smaller increasingly quickly toward low
luminosities. This  
means that if low luminosity sources are concentrated around  
some particular heliocentric distance $r$, these will be missing from the
counts unless the sensitivity is set accordingly. An order of
magnitude in
sensitivity implies a factor of $\sqrt{10}$ in $r_{max-obs}$. Assuming 
masers lie in a ring-like structure around the galactic centre (with
some known radial profile as e.g. a Gaussian), such a
difference in $r_{max-obs}$ could mean either including or not a
factor $\approx$2 sources in the final counts, as $r_{max-obs}$ could
either include only the near  peak or both the near {\it and} the far
peaks of the distribution. 

To better understand the influence of the luminosity function in the
present study we follow the following procedure:

\begin{enumerate}
\item Prescribe a functional form for the luminosity function
  $G(L)$. We require that this is normalised so that $\int
  S(\ln(L))\,\d(\ln(L)) = 1$, where $S(\ln(L))$ is the distribution of
  luminosities sampled in equal logarithmic intervals. From this
  function, we construct a discrete set of 
  probabilities $P_j$, centered on luminosities $L_j$, such that 
  $\sum_j\,P_j(L_j) = 1$. 
\item Set a sensitivity limit $\Phi$ for the observations. For every luminosity
  $L_j$, $\Phi$ sets a maximum heliocentric distance $r_{max-obs}(L_j,\Phi)$ to
  which the observations will detect sources of that particular luminosity;
\item For each luminosity bin we then:
\begin{itemize}
\item Multiply the spatial distribution $F(R)$ (without any
  distance cutoff) with $P(L_j)$, i.e. the probability which
  corresponds to that luminosity. In this way we are left with the
  distribution of sources of luminosity $L_j$
  in the Galaxy;  
\item Apply the distance cutoff $r_{max-obs}(L_j,\Phi)$ and 
  count the number of sources in each galactocentric distance bin
  (this is equivalent to integrating over $\theta$ and $R$ as
  eq.~\ref{eq:ntobs}). In this way we count how many sources of
  the luminosity $L_j$ we are able to detect with the sensitivity $\Phi$.
\end{itemize}
\end{enumerate}

The total number of {\it observed} sources $N_{tot-obs}$ is given by
the sum of all 
(detected) sources at all luminosities:
\begin{equation}
N_{tot-obs} = \int\!\!\!\int\,\left[\sum_{L_j} P(L_j) \times
  F_{j-trunc}(R)\right] \,R\,\dR\,\dt 
\label{eq:ntot}
\end{equation}
where $F_{j-trunc}(R)$ is the spatial distribution of sources within 
$r_{max-obs}(L_j,\Phi)$ and the integration limits of both integrals
are the same as in equation \ref{eq:ntobs}. 

Once the shapes of the spatial distribution $F(R)$ and of the luminosity
function $G(L)$ are chosen, the fit to the histogram in Fig. \ref{fig:n_gc}
has the following free parameters: the total number of sources in the Galaxy
$N$, the parameters defining $G(L)$ and the sensitivity $\Phi$. Note
that we assume that the mean and width of the spatial distribution are
fixed.

To keep the number of free parameters to a minimum, we choose the luminosity
function to be of the simplest kind, a power-law with sharp cutoffs $L_{min}$
and $L_{max}$: 
\begin{equation}
G(L) = A\,L^{\alpha} 
\end{equation}
where $A$ is determined by the normalisation condition, and the only 
free parameter defining $G(L)$ is its power $\alpha$. We take the range of
luminosities from the literature, $L_{min} = 10^{-8}\,L_{\odot}$
and $L_{max} = 10^{-3}\,L_{\odot}$ \citep{wal97}, as our best
estimate.

\subsubsection{Results}
\label{sec:res}

From eq.~\ref{eq:ntot} it is not
possible to determine $N$ and $\alpha$ simultaneously. Nevertheless,
this method is still able to provide  
important insights and predictions on both the total
number of sources in Galaxy ($N$) as well as on the slope $\alpha$ of
the maser luminosity function.  

\begin{figure}
\begin{center}
\includegraphics[width=8cm]{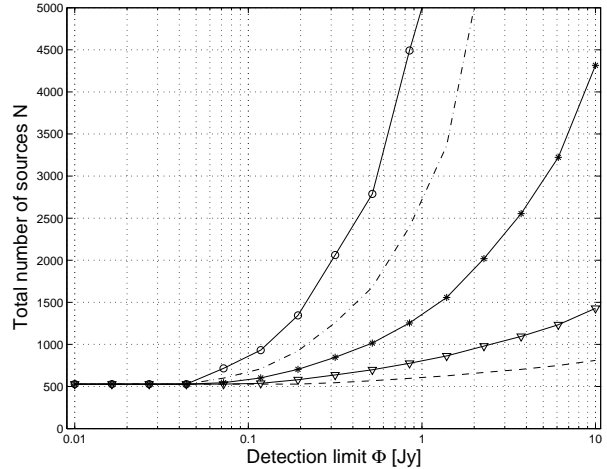}
\caption{$N$ versus detection limit at several values of
  $\alpha$, the slope of the luminosity function (dashed -1; triangles
  -1.2; stars -1.5; dash-dot -2.0; circles -3.5). The spatial
  distribution profile used here is a Gaussian. The detection limit is
  defined as $\Phi$ spread over  
  0.2\,kms$^{-1}$. $N$ is the result of fitting eq.~\ref{eq:ntot}
  to the histogram in Fig. \ref{fig:n_gc}. This explains the
  convergence to $N=519$ at low detection limits. }
\label{fig:res_nsens}
\end{center}
\end{figure}

\begin{figure}
\begin{center}
\includegraphics[width=8.1cm]{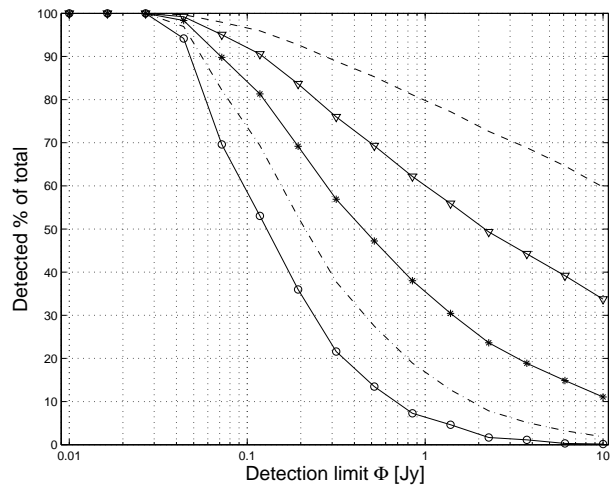}
\caption{$N_{tot-obs}$ as percentage of $N$ versus sensitivity limit at
  different values of $\alpha$, the
  slope of the luminosity function. The curves are labeled as in
  Fig. \ref{fig:res_nsens}. Note that an order of magnitude   
  in sensitivity implies a factor of $\sqrt{10}$ in 
  $r_{max-obs}$.  The spatial distribution profile used here is a
  Gaussian. The detection limit is defined as $\Phi$ spread over
  0.2\,kms$^{-1}$.} 
\label{fig:res_percentage}
\end{center}
\end{figure}

We can attempt to answer two different questions: 
\begin{itemize}
\item Knowing the mean detection limit in GCMM and how many sources
  GCMM contains, how many sources
  should there be in total in the Galaxy and how does this number
  depend on the slope of the luminosity function?
\item What fraction of $N$ do we expect to detect depending on the
  detection limit and the slope of the luminosity function?
\end{itemize}

The answer to the first of these question is shown in
Fig. \ref{fig:res_nsens}. The 
curves in the graph were obtained by fitting equation \ref{eq:ntot} to
the histogram in Fig.~\ref{fig:n_gc}, excluding the
integral over $R$. For example, if we assume the mean detection
limit of GCMM to be 1\,Jy, $N$ would lie between $\approx 520$ and $\approx
5000$, depending on the power of the luminosity function. If we are to
constrain our range for $N$ by taking into
account the prediction made in Sect.~\ref{sec:spadistr} ($N \approx
1000$) and of \citealt{vdw05} ($N \approx 800-1200$), we can  
state that  the luminosity function has a slope between $\approx -1.2$
and $\approx -1.5$. At the detection limit of the MMB Survey (0.5\,Jy)
the total number of sources is expected to lie between 519 and more
than 2500, depending on the slope of the luminosity function. 

The answer to the second question is shown in
Fig.~\ref{fig:res_percentage}. In this approach the absolute number of sources
$N$ and $N_{tot-obs}$ are irrelevant,
and only the shapes of the spatial and luminosity distributions are
important.  From the graph we can say that by taking the mean sensitivity
of GCMM to be 1\,Jy, the total number of sources in that catalogue
(519) represents  
between 10 and 80\% of the total depending on the chosen luminosity
function. Again, if previous estimates of the total number of methanol
masers in the Galaxy are correct, we can state that GCMM contains
$30-50\%$ of $N$, and this would constrain the slope of the luminosity
function $\alpha$ to be $> -1.5$. The detection limit of the MMB
Survey (0.5\,Jy) will produce a catalogue containing between 15 and
85\% of the real total number of sources in the Galaxy, depending on
the slope of the luminosity function. 

Fig. \ref{fig:res_percentage} indicates that deeper
observations have the effect of reducing the dependence of the
fraction $N_{tot-obs}/N$ from the luminosity function. A detection
limit of 0.1\,Jy would allow to detect $60-95\%$ of the total number
of sources, which is a more favourable range than the one provided by
a mean detection limit of 1\,Jy. Assuming a power-law with slope
$-1.5$, a 0.1\,Jy detection limit would yield the detection of 85\% of the
total number of existing methanol masers in the Milky Way.

\section{Can we constrain the luminosity function from GCMM?}
\label{sec:lumin}

\begin{figure}[!ht]
\includegraphics[width=9cm]{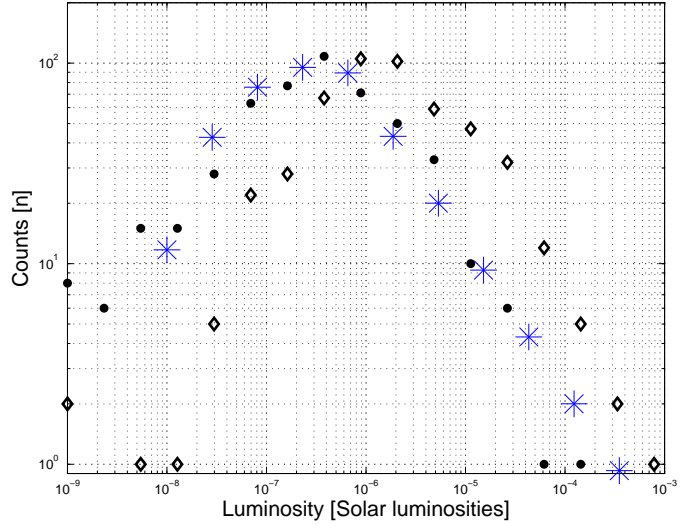}
\caption{Distribution of the luminosities of all masers in GCMM,
  assuming, for all ambiguous sources, either the near (dots) or the
  far (diamonds) heliocentric distance. The luminosities were
  calculated by assuming that the peak
  flux is spread over 0.2\,kms$^{-1}$.  The difference in the
  peak position is of about 1 order of magnitude. Stars show the
  results obtained by the observation at 2\,Jy sensitivity of the
  modelled maser population in the Galaxy. The model is defined by a
  total number of sources equal to 5000, a spatial distribution as in
  Fig. \ref{fig:n_gc} and a luminosity function expressed as a single
  power-law of index -1.7 between $10^{-8}$ and $10^{-3} L_{\odot}$ .}
\label{fig:lum_distr}
\end{figure}

Fig. \ref{fig:lum_distr} shows the distribution of luminosities of the
masers in GCMM. Dots and diamonds represent the two extreme cases where
all {\it ambiguous} sources have been assumed to lie at the near and far
heliocentric distance, respectively. Since the sources suffering from
distance ambiguity are about 90\% of the catalogue, the two curves in
Fig.\ref{fig:lum_distr} show a relative shift of about one order of
magnitude in luminosity.

As GCMM is a highly heterogeneous sample (different
sensitivities, non-uniform area coverage), we cannot conclude that the
plots in Fig.\ref{fig:lum_distr} are indicative of the shape of the
{\it real} luminosity 
function of the masers. Nevertheless, if GCMM does not contain most of
the faint methanol masers in the Galaxy ($< 1-3$\,Jy), we expect the
counts of a large scale deep search of methanol masers in the Milky
Way (as e.g. the MMB Survey) to populate the lower part of the
luminosity range in  
Fig.~\ref{fig:lum_distr}. This is due to the fact that a 1\,Jy source
is intrinsically fainter than the peak luminosity in those
distribution, as calculated in Sect.~\ref{sec:howmany}. The
detection of many sources at the 1\,Jy level or below will change the
shape of the luminosity distribution in Fig.\ref{fig:lum_distr} closer
to one of a power-law. 

Maintaining the assumption of the luminosity function of methanol
masers to be a single power-law between sharp cutoffs and observing it
with a sensitivity of e.g. 2\,Jy we obtain best similarity to the data
as shown in Fig.\ref{fig:lum_distr}. The synthetic data
points (stars) seem closest to the data curve obtained when putting all
ambiguous sources at the near heliocentric distance (dots). This
strongly suggests that most of the ambiguous sources in GCMM probably
lie at the near heliocentric distance. It is to notice that the 
last points toward lower luminosities could be considered to be the
only few sources at the far distance, what would bring them within the
range of luminosities assumed in the model ($10^{-8} < {\rm L} <
10^{-3}\, L_{\odot}$).

It is in principle possible to eliminate the near-far distance problem
by selecting appropriate subsamples of sources from GCMM. We have then
two possibilities: either we select the sources in the outer Galaxy
($90^{\circ} \le l \le 270^{\circ}$) or we select all sources at the
tangent point of the molecular ring ($\pm 50^{\circ} \le l \le \pm
20^{\circ}$). The luminosity distributions resulting from these selections
were not better defined than the ones shown in
Fig. \ref{fig:lum_distr}. This is mainly due to the fact that the
number of sources we are left with after the selection ($\approx 90$)
is severely reduced as compared to the total. Also, the selection of
sources at the tangent points does not completely eliminate the
near-far ambiguity, because of the difficulty of unambiguously
defining the tangent point.  

It is important to notice that our estimates of the index of the
luminosity function of methanol masers ($-1.5 \ge \alpha \ge -2$) is in slight
disagreement with what found in previous studies ($\approx
-2$). The difference in
index could probably be reduced assuming a more complex luminosity
function, as e.g. a broken power-law. We still consider the luminosity
function presented here to be the best estimate with the least number
of assumptions and free parameters, and, most importantly, obtained
without any selection of sources. The definitive shape of the
luminosity function of methanol masers in the Galaxy will be found
once the MMB Survey is completed.


\section{Conclusions}
\label{sec:concl}

Using the General Catalogue of 6.7\,GHz Methanol Masers in the Milky Way, we
have modelled their spatial distribution in the Galaxy and estimated their
luminosity function. We conclude the following: 

\begin{itemize}
\item methanol masers are distributed in a ring of some 5\,kpc
  in radius around the Galactic centre;
\item most of the sources in GCMM are probably confined in a ring of
  heliocentric radius equal to $R_0$ or slightly larger. This also
  means that the sources showing heliocentric distance ambiguity lie
  probably at the near distance;
\item the luminosity function of methanol masers in our Galaxy is modelled
  as a power-law between sharp cutoffs having an index $\alpha$
  between $-1.5$ and $-2$;  
\item given a uniform survey which is able to define a methanol maser
  luminosity function, the approach presented here will allow us to
  estimate the 
  total number of sources in the Galaxy. As an example, a large scale
  survey with a sensitivity limit of 0.5\,Jy will be able to detect
  $\approx 50\%$ of the total number of sources or more, if the power
  of the luminosity function is $-1.5$ or lower. This is what is
  expected from the MMB Survey. 
\end{itemize}


\begin{acknowledgements}
We thank M. Thompson for his comments on the manuscript that greatly improved
its understandability. M.P. thanks A. Pedlar for the discussions that
in the very early days of this paper were very helpful in defining the
starting point of the presented considerations. 
\end{acknowledgements}

\bibliography{methanol,varia,othermasers,starformation+IR,surveys_cat,tech}
\bibliographystyle{aa}

\end{document}